\input{aipcheck}

\documentclass[final]{aipproc}

\usepackage{slashed}
\usepackage{amsmath,amssymb,amsfonts,amsbsy}
\usepackage{commath}

\layoutstyle{6x9}

\begin{document}

\title{Higgs boson photoproduction at the LHC}


\author{M.B. Gay Ducati and G.G. Silveira}{
address={High Energy Physics Phenomenology Group, UFRGS, \\ Caixa Postal 15051, CEP 91501-970 - Porto Alegre, RS, Brazil.}
}

\begin{abstract}We present the current development of the photoproduction approach for the Higgs boson with its application to $pp$ and $pA$ collisions at the LHC. We perform a different analysis for the Gap Survival Probability, where we consider a probability of 3\% and also a more optimistic value of 10\% based on the HERA data for dijet production. As a result, the cross section for the exclusive Higgs boson production is about 2 fb and 6 fb in $pp$ collisions and 617 and 2056 fb for $p$Pb collisions, considering the gap survival factor of 3\% and 10\%, respectively.
\end{abstract}

\maketitle

\section{INTRODUCTION}
\label{sec:intro}

As an alternative way to study the Higgs boson production at the LHC, the Central Exclusive Diffractive (CED) production has been currently analyzed as a new framework for particle production \cite{standard-candle}. Indeed, the Double Pomeron Exchange (DPE) and the two-photon process offer the opportunity to study the exclusive Higgs boson production in proton and nuclei collisions. One way to increase these cross sections is to consider nuclei collisions, especially for the two-photon process, where the photon flux is enhanced by a factor of $Z$ in $pA$ collisions, and $Z^{2}$ in $AA$ ones. For instance, the predicted cross section for the Higgs boson production in PbPb collisions is 18 pb \cite{denterria}, which is enhanced by five orders if compared to the predictions for $pp$ collisions (0.18 fb). However, in DPE this enhancement is smaller, showing an increasing from 3 fb for $pp$ collisions to 100 fb in AuAu ones \cite{levin-miler}. In this sense, we compute the cross sections with the photoproduction mechanism for the CED Higgs boson production in hadron-hadron and hadron-nucleus collisions at the LHC.

\section{PHOTOPRODUCTION MECHANISM}\label{photo}

Considering the $\gamma p$ interaction in high-energy collisions, we compute the cross section for the CED Higgs boson production by DPE in the $\gamma p$ subprocess \cite{peiHiggs}, which is one of the possible subprocess in Ultraperipheral Collisions (UPC) \cite{hencken}. The photon fluctuates into a quark-antiquark pair, and then the interaction occurs between the proton and this pair by the exchange of gluons in the $t$-channel.

\begin{figure}
\includegraphics*[scale=0.55]{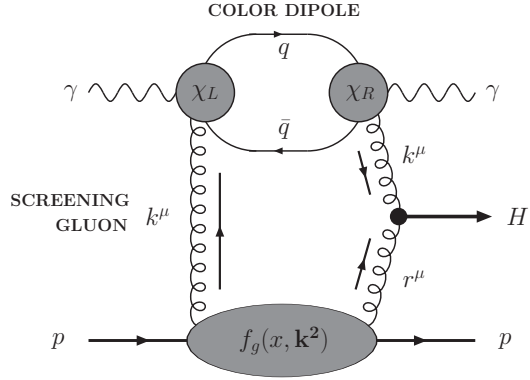}
\caption{\label{fig1}Diagram that represents the photoproduction mechanism for the Higgs boson production.}
\end{figure}

The diagram at partonic level is taken into account in order to compute the scattering amplitude, where a quasi-real photon interacts with a quark nto the proton. The imaginary part of the scattering amplitude is computed by the use of the Cutkosky Rules ${\rm{Im}}A = \frac{1}{2} \int \dif(PS)_{3} \, A_{L} \, A_{R}$, with $A_{L}$ and $A_{R}$ being the amplitudes in the left- and the right-hand sides of the central line that splits the diagram in Fig.\ref{fig1} in two pieces, and $\dif (PS)_{3}$ is the volume element of the three-body phase space. This integration results in the following amplitude
\begin{eqnarray}
\mbox{Im} A_{T} = - \frac{s}{6} \frac{M^{2}_{H}}{\pi v}\frac{\alpha_{s}}{N_{c}} \int \Phi^{T}_{\gamma\gamma}(\mathbf{k}^{2},Q^{2}) \frac{\dif\mathbf{k}^{2}}{\mathbf{k}^{2}},
\label{amp-im}
\end{eqnarray}
where $\Phi^{T}_{\gamma\gamma}$ is the impact factor for the $\gamma$-$\gamma$ transition
\begin{eqnarray}
\Phi^{T}_{\gamma\gamma}(\mathbf{k}^{2},Q^{2}) = 4 \pi \alpha \alpha_{s} \sum_{q} e^{2}_{q} \int^{1}_{0} \dif\tau \dif\rho \frac{\mathbf{k}^{2}[\tau^{2} + (1 - \tau)^{2}][\rho^{2} + (1 - \rho)^{2}]}{Q^{2}\rho(1 - \rho) + \mathbf{k}^{2}\tau(1 - \tau)},
\label{imp-fact}
\end{eqnarray}
$Q^{2}$ is the virtuality of the initial photon, $v$ = 246 GeV is the vacuum expectation value of the Electroweak Theory, $e_{q}$ is the charge of the quark in the dipole, $\alpha$ and $\alpha_{s}$ are the electromagnetic and strong coupling constants, respectively, $\tau$ is the Feynman parameter, and $\mathbf{k}$ is the transverse momentum of the gluons. In this calculation, we introduce the Sudakov parametrization $k^{\mu} = \rho q^{\prime \mu} + \beta p^{\mu} + k^{\mu}_{\perp}$, with $q^{\prime\mu} = q^{\mu} - \frac{Q^{2}}{s}p^{\mu}$.

In order to include all the partonic content of the proton, one has to replace the contribution of the $gq$ coupling by the unintegrated parton distribution function $f_{g}(x,\mathbf{k}^{2},\mu^{2}) = K \, \partial G(x,\mathbf{k}^{2})/\partial \ln \mathbf{k}^{2}$, where the function $G(x,\mathbf{k}^{2})$ is the integrated gluon distribution function, and the multiplicative factor $K$ = 1.2 takes into account the non-diagonality of the distribution \cite{shuvaev}. In this work we apply the MSTW2008LO parametrization for such distribution function \cite{mstw2008}.

To obtain the event rate, one have to integrate the amplitude squared given by Eq.(\ref{amp-im}) over the momenta of the particles in the final state, including the prescription for $f_{g}$. The result for central rapidity reads
\begin{eqnarray}
\left. \frac{\dif \sigma}{\dif y_{H}\dif t} \right|_{y_{H},t=0} = S^{2}_{gap} \frac{K_{NLO}\alpha^{4}_{s}}{288 \pi^{5}B}\left( \frac{M^{2}_{H}}{N_{c} v} \right)^{2} \left[ \int^{\mu^{2}}_{\mathbf{k}^{2}_{0}} \frac{\dif \mathbf{k}^{2}}{\mathbf{k}^{2}} \tilde{f}_{g}(x,\mathbf{k}^{2},\mu^{2}) \Phi^{T}_{\gamma\gamma}(\mathbf{k}^{2},Q^{2}) \right]^{2},
\label{gammap-xsec}
\end{eqnarray}
where $K_{NLO}$ is taken to 1.5, which corresponds to the enhancement of the $gg \to H$ cross section at NLO accuracy \cite{spira}, and $B$ = 5.5 GeV$^{-2}$ is the slope of the gluon-proton form factor. The function $\tilde{f}(x,\mathbf{k}^{2},\mu^{2}) = \sqrt{T(\mathbf{k}^{2},\mu^{2})}G(x,\mathbf{k}^{2})$ is the modified unintegrated gluon distribution function that includes the Sudakov form factor $T$ computed at Leading Logarithm Accuracy (LLA).

Regarding the phenomenological aspects introduced in this result, the Rapidity Gap Survival Probability (GSP) depends particularly of the process under consideration. The GSP for the $\gamma p$ process is not computed yet, and we use the one of 3\% predicted for the Pomeron-Pomeron mechanism. However, we expect a higher survival factor for the $\gamma p$ interaction, since the large distances between the two colliding hadrons in UPC should decrease the probability of interaction between secondary particles. Analyzing the results for central dijet production at HERA, one finds that the survival probability is about 10\% \cite{kaidalov-hera}, and we make predictions with this probability for the CED Higgs boson photoproduction.

\section{ULTRAPERIPHERAL COLLISIONS}\label{upc}

The initial photon is emitted from one relativistic source object, which can be a proton or a nucleus. Particularly, a nucleus has $Z$ protons, which enhances the photon flux in $pA$ and $AA$ collisions. In fact, considering the luminosity and pile-up effects in the collisions at the LHC, the $pA$ collisions may offer the best experimental condition if compared to $pp$ and $AA$ collisions \cite{denterria}. Additionally, in the photoproduction mechanism, we neglect the contribution from $AA$ collisions, since the shadowing effects present in the nuclear PDF will decrease the production cross section by a factor of 0.2-0.3. The production cross section in UPC is given by
\begin{eqnarray}
\sigma_{had} = 2 \int_{\omega_{min}}^{\omega_{max}} \frac{\dif n_{i}}{\dif\omega} \sigma_{\gamma p},
\label{had-xsec}
\end{eqnarray}
where $\omega_{min} = M^{2}/2x\sqrt{s_{NN}}$ and $\omega_{max} = \sqrt{Q^{2}\gamma^{2}_{L}\beta^{2}_{L}}$, and $\sigma_{\gamma p}$ is given by Eq.(\ref{gammap-xsec}). The functions $\dif n_{i}/\dif\omega$ are the photon fluxes for protons and nuclei, which can be found in Ref.\cite{ppHiggs}. In this sense, the photon virtuality is decomposed into $Q^{2} = -\omega^{2}/\gamma^{2}_{L}\beta^{2}_{L} - \mathbf{q}^{2} \leq R^{-2}_{i}$, with $\gamma_{L} = (1-\beta_{L}^{2})^{-1/2} = \sqrt{s}/2m_{p}$, which is restricted by the coherence condition in UPC, depending of the radius of the source object.

\section{RESULTS}\label{res}

The hadronic cross section is computed for $pp$, $p$Pb, $p$Au, $p$Ar, and $p$O collisions at the LHC. Actually, collisions involving gold nucleus are not going to be measured in the LHC, however we include such prediction to compare with previous results \cite{levin-miler}. The Tab.\ref{tab1} shows the kinematics introduced in this calculation, and the predicted cross sections for the CED Higgs boson photoproduction for the two possibilities of the survival factor.

\begin{table}[h!]
\begin{tabular}{cccccc}\hline
              & $\sqrt{s_{NN}}$ (TeV) & $\gamma_{L}$ & R (fm) & \multicolumn{2}{c}{$\sigma_{had}$ (fb)} \\ \hline 
$S^{2}_{gap}$ &                       &              &        & 3\%      & 10\%                         \\ \hline
$pp$          & 14.0                  & 7460         & 0.7    & 1.77     & 5.92                         \\ \hline
$pO$          & 9.90                  & 5314         & 3.0    & 2.31     & 7.70                         \\ \hline
$pAr$         & 9.40                  & 5045         & 4.1    & 21.56    & 71.87                        \\ \hline
$pAu$         & 8.86                  & 4755         & 7.0    & 768.     & 2562.                        \\ \hline
$pPb$         & 8.80                  & 4724         & 7.1    & 617.     & 2056.                        \\ \hline
\end{tabular}
\caption{\label{tab1}The predicted cross sections for the CED Higgs boson photoproduction at the LHC for $M_{H}$=120 GeV, and the kinematics parameters. The cross section is shown for the two possibilities of the GSP: 3\% and 10\%.}
\end{table}

As one can see, the production cross section is significantly enhanced in $pA$ collisions taking nucleus with high $Z$. These results are higher than the ones obtained for the two-photon and for the DPE mechanism. In the case of $pp$ collisions, the cross section is similar to that of the DPE mechanism, but one order higher than that for the two-photon mechanism. Considering the $AA$ run that will occur in the end of 2010, new data can be available for nuclei collisions in the next year.

\section{CONCLUSIONS}

In this work we applied the photoproduction mechanism for the CED Higgs boson production to $pp$ and $pA$ collisions in the LHC. The results show an  enhanced cross sections for collisions involving Au and Pb nucleus, which open a new way to detect the Higgs boson in the LHC. The GSP is a fundamental aspect to be determined with the future data from the LHC, playing an important role for reliable predictions of diffractive processes. Therefore, the photoproduction mechanism offers a new approach for the Higgs boson production, showing a cross section competitive with other production mechanisms.

\begin{theacknowledgments}
This work was partially supported by CNPq.
\end{theacknowledgments}

\end{document}